\documentclass[aps,prb,twocolumn,groupedaddress,showkeys,showpacs]{revtex4}
\usepackage{multirow,amssymb,amsbsy,amsmath}
\usepackage{dcolumn}% Align table columns on decimal point
\usepackage{bm}% bold math
\usepackage{graphicx}
\usepackage{xspace}

\newcommand{\scl}{$\mathrm{(Sr,Ca,La)_{14}Cu_{24}O_{41}}$}
\newcommand{\sr}{$\mathrm{Sr_{14}Cu_{24}O_{41}}$}

\newcommand{\nacu}{$\mathrm{Na_{1+x}CuO_{2}}$}
\newcommand{\cay}{$\mathrm{Ca_{2+x}Y_{2-x}Cu_5O_{10}}$}

\newcommand{\mb}{$\mu_B$~}
\newcommand{\op}[1]{%
    \fontdimen12\textfont3=2pt\fontdimen12\scriptfont3=1.4pt%
    \!\null\mathop{\vphantom{#1}\smash{#1}}\limits_{\sim}\null\!}
\newcommand{\xref}[1]{\protect\ref{#1}}
\newcommand{\figref}[1]{Fig.~\protect\ref{#1}}
\newcommand{\fmref}[1]{(\protect\ref{#1})}
\begin{document}

\title{Revisiting the chain magnetism in Sr$_{14}$Cu$_{24}$O$_{41}$: Experimental and numerical results}
\author{R. Klingeler$^{1,2}$}
\email{r.klingeler@ifw-dresden.de}
\author{B. B\"{u}chner$^1$}
\author{K.-Y. Choi$^3$}
\author{V. Kataev$^1$}
\email{v.kataev@ifw-dresden.de}
\affiliation{$^1$Leibniz-Institute for Solid State and Materials Research
IFW Dresden, 01171 Dresden, Germany}
\affiliation{$^2$Laboratoire National des
Champs Magn\'{e}tiques Puls\'{e}s, 31432 Toulouse, France}
\affiliation{$^3$Institute for
Materials Research, Tohoku University, Katahira 2-1-1, Sendai 980-8577,
Japan}
\author{U. Ammerahl}\author{A. Revcolevschi}
\affiliation{Laboratoire de Physico-Chimie des Solides, Universit\'e Paris-Sud, 91405 Orsay
C\'edex, France}
\author{J. Schnack}
\affiliation{Fachbereich Physik, Universit\"at Osnabr\"uck, D-49069 Osnabr\"uck}

\date{\today}

\begin{abstract}

  We study the magnetism of the hole doped CuO$_2$ spin chains
  in \sr\ by measuring the Electron Spin Resonance (ESR) and the
  static magnetization $M$ in applied magnetic fields up to
  14~T. In this compound, the dimerized ground state and the
  charge order in the chains are well established. Our
  experimental data suggest that at low temperatures the
  Curie-like increase of $M$ as well as the occurrence of the
  related ESR signal are due to a small amount of paramagnetic
  centers which are not extrinsic defects but rather unpaired Cu
  spins in the chain. These observations qualitatively confirm
  recent \emph{ab initio} calculations of the ground state
  properties of the CuO$_2$ chains in \sr. Our complementary
  quantum statistical simulations yield that the temperature and
  field dependence of the magnetization can be well described by
  an effective Heisenberg model in which the ground state
  configuration is composed of spin dimers, trimers, and
  monomers.

\end{abstract}


\maketitle

\section{INTRODUCTION}

Hole doped antiferromagnets with an inhomogeneous ground state
have attracted much attention in recent years. A prominent
example is provided by high temperature (high-T$_c$)
superconductors in which low dimensionality, $s=1/2$ quantum
magnetism, and charge degrees of freedom are intimately related.
In particular, a nonuniform distribution of charges in the Cu-O
planes of the high-T$_c$ layered cuprates seems to play an
important role in establishing the
superconductivity.\cite{KBF:RMP03} The quasi one-dimensional
(1D) edge-sharing CuO$_2$ chain compounds such as \scl , \nacu , or
\cay\ constitute another class of cuprates with charge
inhomogeneities.\cite{UNA:JPSJ96,Dag:RPP99,HSM:PRL05,MKK:PRB05} The
interplay of spin and charge degrees of freedom in one
dimension, as is evidenced by both theoretical and experimental
results, is even more pronounced as compared to the layered 2D
situation, yielding unusual ground states and novel excitations
both in the spin and charge sectors.

In this work we focus on the low-temperature magnetic properties of the so-called ``telephone
number'' compound \sr, where a composite structure with Cu$_2$O$_3$ spin ladders and CuO$_2$ spin
chains is realized. Magnetically, the ladders do not contribute significantly at low temperature
due to a large spin gap of $\Delta \sim 380$\,K. \cite{TKK:JPSJ96,EUA:PRL98,KNA:PRL99,HBA:PRB01} A remarkable
property of this compound is self-doping, i.e. the material is intrinsically hole doped with 25\%
of holes per Cu site. As most of the holes are localized in the chains their concentration in this
subunit is very high, close to 60\% per Cu in the chain formula unit (c.f.u.). This influences
strongly the magnetism of the spin chain resulting in an inhomogeneous ground state. Two Cu spins
which are separated by a hole site couple antiferromagnetically (AF). If they are surrounded by
two localized holes at both sides they form almost independent dimers. Such a dimerized
nonmagnetic state in the chain subsystem of \sr\ is experimentally well established.
\cite{KSK:PC96,EAT:PRB96,MKY:PRB96,RBM:PRB99,MYK:PRB99} The occurrence of charge order, which is an
important ingredient of the dimer scenario in \sr , is supported e.g. by NMR and ESR data.
\cite{TME:PRB98,KCG:PRB01} In a very recent numerical study\cite{GeL:PRL04} Gelle and Lepetit (GL)
highlight that structural modulations due to the pseudoperiodicity of the chains and the ladders
might be crucial for the electron localization and for the occurrence of the dimer state.
Remarkably, these calculations suggest that the ground state is not perfectly dimerized even in
the limiting case where all holes are located in the chains. Ideally, in that case, where six out
of the ten sites in the chain unit cell are holes, the remaining four spins could be coupled to
two AF dimers with no free spins left. In contrast to that, GL arrive at the surprising result
that it is energetically favorable if a few spins are left unpaired. Then the stable configuration
corresponds to 1.7 dimers and 0.5 free spins per c.f.u., respectively. Therefore, the occurrence
of a Curie contribution to the magnetic susceptibility of \sr\ at low temperatures, which has
originally been attributed to the free spins, might be an intrinsic property of the hole doped
Heisenberg Cu-O spin chain due to the distortion by structural modulations.

The aim of our paper is thus twofold. First, we show
experimental magnetization and electron spin resonance (ESR)
data which give strong indications that a small amount of free
spins, that gives rise to a finite magnetization and to an ESR
signal at low temperatures, is not extrinsic but mainly reside
in the chain subunit of \sr . This finding {\em qualitatively}
supports the model calculations by GL in
Ref.~\onlinecite{GeL:PRL04}.  Secondly, we present our own quantum
statistical simulations which rely on the numerical
diagonalization of an effective Heisenberg
Hamiltonian.\cite{ScO:JMMM05,Sch:EPJB05} This Hamiltonian, which
depends parametrically on hole positions and on the screened
Coulomb interaction between them, also yields an inhomogeneous
ground state configuration of spins and holes. It turns out that
the ground state is dominantly built of the aforementioned
dimers, but also contains weakly coupled monomers and/or trimers
which not only leads to the Curie-like behavior of the
magnetization at low temperatures, but also yields a nearly
perfect \emph{quantitative} description of the low-temperature
magnetic response of the Cu spin chain. The very good agreement
also suggests that other arrangements of spins and holes are
well gaped from the ground state configuration.

\section{Experimental results}

In this section we present ESR and magnetization measurements on
a high quality single crystal of \sr . For ESR experiments an
X-Band Bruker ESR spectrometer has been used, which operates at
a frequency of 9.5~GHz. The magnetization data have been
collected with a homemade vibrating sample magnetometer (VSM) in
external magnetic fields up to 14~T and with a SQUID
magnetometer in a field of 1~T. The single crystal grown by the
floating zone technique\cite{AmR:JCG99} has been previously
thoroughly characterized by measuring the magnetization, the
thermal expansion,\cite{ABC:PRB00} the electrical and thermal
transport,\cite{HBA:PRB01,HEB:PRL04} the inelastic neutron
scattering cross section,\cite{RBM:PRB99}, and the ESR
response.\cite{KCG:PRB01}

%===================    figure   =================================
\begin{figure}
\center{\includegraphics [width=0.8\columnwidth,clip]{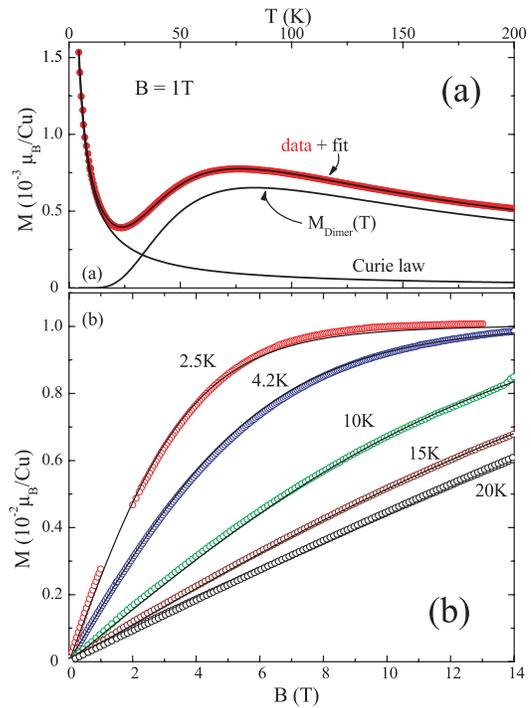}} \caption[]{\label{sr14}(color online) Magnetization of \sr\
  versus temperature for a constant magnetic field of $B=1$~T
  (a) and versus applied magnetic field for $2.5$~K$\leq T \leq$
  $20$~K (b). $B$ was applied along the $c$-axis. Lines are fits
  to the data (see text).}
\end{figure}
%===================    figure   =================================

In order to recall the magnetic properties of \sr\ we show in
\figref{sr14}(a) the temperature dependence of the
magnetization $M(T)$ of our single crystal in a magnetic field
of $B=1$~T applied parallel to the crystallographic $c$-axis
(chain direction). The $M(T)$ dependence can be very well fitted
with Eq.~\fmref{dimer} comprising two terms, the magnetization
$M_{dimer}$ of AF coupled dimers in a concentration $N_D$ with
an exchange constant $J/k_B = -134$~K, and the Curie
magnetization $M_{Curie}$ owing to a small concentration $N_S$
of free spins (see, e.g. Ref.~\onlinecite{KCG:PRB01})

\begin{eqnarray}
\label{dimer}
M(T) &=& M_{dimer} + M_{Curie}
\\
&=&
\frac{N_Ag^2\mu_B^2B}{24k_B}\left[\frac{2N_D}{T[3 +
    \exp(-J/k_BT)]}+\frac{N_S}{4T}\right]
\ .
\nonumber
\end{eqnarray}

The very good agreement of the experimental $M(T)$ curve with
Eq.~\fmref{dimer} give evidence of the dimerized ground state
which in addition has been proven for our sample by inelastic
neutron scattering and thermal expansion
measurements.\cite{RBM:PRB99,ABC:PRB00} In \figref{sr14}(b) we
present the magnetization versus applied magnetic field, $M(B)$,
at temperatures of 20~K, 15~K, 10~K, 4.2~K, and 2.5~K,
respectively.  At all temperatures the dependence of $M$ on $B$
can be nicely fitted as a sum of the temperature independent
contribution $\chi_0\cdot B$ owing to diamagnetism and Van-Vleck
paramagnetism of the Cu ions and a Brillouin function
$B_{s=\frac{1}{2}}$ describing the magnetization of free $s=1/2$
spins with a concentration $N_S$

\begin{equation}
M(B) = \chi_{0} \cdot B + \frac{1}{2}N_Sg_c\mu_B \cdot
B_{s=\frac{1}{2}}\left(\frac{g\mu_B(B+\lambda M)}
{2k_BT}\right)
\ .
\label{Brill}
\end{equation}

With a $g$ factor of $g_c = 2.04$ (see below) one obtains
$\chi_{0} \approx 1\cdot 10^{-5}$\,emu/Mol Cu, the mean field
parameter $\lambda =0$ and $N_S \simeq 0.01$/Cu. We note that
the magnetization at $T = 2.5$~K saturates in a field of
$B=14$~T at a very small value of about $1\cdot 10^{-2}$\,\mb
/Cu justifying that the magnetic response at low temperature is
caused by about 1\% of free $s = 1/2$ spins while there is no
response from the remaining 99\% of the spins. Moreover, we
emphasize that the determination of $N_S$ from our high field
magnetization data is much more straightforward and accurate as
compared to the conventional analysis of the low temperature
contribution to $M(T)$ using the $M_{Curie}$ term in
Eq.~\fmref{dimer}. Therefore, precise knowledge of $N_S$ from
the analysis of $M(B)$ curves allows for an accurate
determination of the number of dimers $N_D$. Substituting
$N_S\,=\,0.01$/Cu in the $M_{Curie}$ term in Eq.~\fmref{dimer}
one obtains from the fit to the experimental $M(T)$ dependence
$N_D = 0.0738$/Cu. Thus the number of magnetic Cu sites amounts
to $N_S + 2N_D \approx 3.78$/f.u. in our \sr\ single crystal
which differs from a previous less precise estimate that
employed the analysis of $M(T)$ curves of polycrystalline
samples.\cite{CBC:PRL96}

We note that the concentrations of dimers and free spins, $N_D$
and $N_S$, respectively, are significant observables in the
recent \emph{ab initio} calculations of the low energy
electronic properties of the Cu-O chain in \sr\ by
GL.~\cite{GeL:PRL04} Therefore, the accurate determination of
$N_D$ and $N_S$ from our magnetization data provides an
important check for the relevance of the GL model with respect
to the low energy physics of \sr .  Comparison of our data with
the numerical results shows that the number of dimers $N_D$
found in our analysis roughly agrees with the calculations of GL
for the case that all holes are localized in the chains,
$N_D^{GL} = 0.0708$/Cu . The number of free spins $N_S$,
however, is smaller than theoretically predicted $N_S^{GL} =
0.021$/Cu by a factor of two. This is surprising because the
experimental value $N_S$ comprises not only the response owing
to the imperfect dimerization of the spin chain, but also should
include the response of paramagnetic defects which may occur in
real crystals.  Thus, theoretical calculations should yield a
smaller number of free spins than experimental data. Besides the
quantitative discrepancy this comparison suggests that, at least
in the framework of the GL model, the free spins in a
concentration $N_S \simeq 0.01$/Cu found in our \sr\ single
crystal are \emph{not extrinsic} paramagnetic defects but mainly
reside in the spin chain. \cite{comment} This suggestion is corroborated with
the analysis of the anisotropy of the ESR signal and of the
magnetization.

\begin{figure}
\center{\includegraphics [width=0.8\columnwidth,clip]{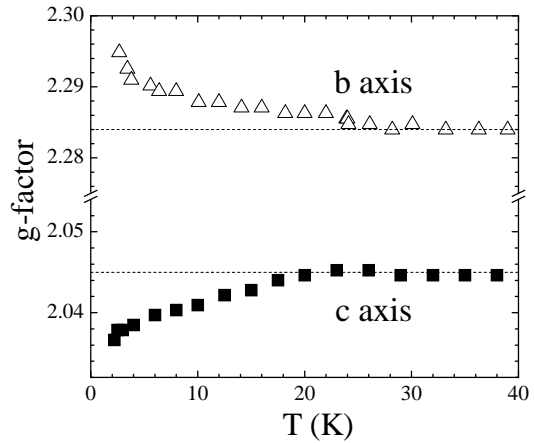}} \caption[] {\label{gfactor}Temperature dependence of the
  $g$-factor along $c$- and $b$-axis, respectively, as obtained
  from ESR data. Dashed lines extrapolate $g$ of the dimers.
  Below $\sim$20\,K, the response is due to small deviations
  from the perfectly dimerized state.}
\end{figure}

Detailed studies of the ESR response from the spin chains in the family of
the ``telephone number'' compounds have been published in
Refs.~\onlinecite{KCG:PRB01,KCG:PRL01}. Here we present the ESR data of
the \sr\ single crystal in the low temperature regime. Below 20~K the ESR
signal is mainly due to the free spins, which also contribute to the
Curie-like increase of the magnetization, compare \figref{sr14}(a). At
higher temperatures the ESR response is dominated by the dimer spins which
are thermally activated. The evolution of the $g$ factor of the ESR signal
in the crossover regime around 20~K is shown in \figref{gfactor}. The $g$
factor of the Cu spins in a dimer is anisotropic with the values
$g_c\,=\,2.045$ for the magnetic field $\vec{B}$ parallel to the chain
direction ( $c$-axis) and $g_b\,=\,2.284$ for the magnetic field $\vec{B}$
perpendicular to the plane of the chains ($b$-axis), respectively. These
$g$ factors are temperature independent. The anisotropy of the $g$ tensor
is determined by the crystal field splitting of the Cu orbital state of
$e_g$ and $t_{2g}$ symmetry

\begin{equation}
g_{c} = 2 - \frac{8\cdot\lambda}{\Delta_{yz,zx}}, g_{b} = 2 -
\frac{2\cdot\lambda}{\Delta_{xy}}
\ .
\label{eqn.g}
\end{equation}

Here $\lambda\,=-0.1$~eV is the spin-orbit coupling constant and
$\Delta_{yz,zx}$ and $\Delta_{xy}$ are the energy differences
between relevant orbital states.\cite{AbB:DP86} The anisotropy
of the $g$ tensor above $\sim\,20$~K is temperature independent
and is typical for a Cu ion in an approximately square planar
coordination of the oxygen ligands,\cite{AbB:DP86,KCG:PRB01}
i.e. the coordination occurring in the chain of \sr.
Remarkably, below 20~K the $g$ factor for both orientations
changes only slightly giving strong indication that the ESR
response at low $T$s comes from a paramagnetic Cu ion having a
very similar environment to that of the chain Cu ions. A small
change in the anisotropy of the $g$ factors can be explained by
a slightly different charge distribution around a free Cu spin
which has in the chain two neighboring holes, both to the left
and to the right hand side, whereas a Cu spin in a dimer has only one
neighboring hole on the one side and two holes on the other
side, compare also \figref{F-1}.

%===================    figure   =================================
%===================    figure   =================================

\begin{figure}
\center{\includegraphics [width=0.8\columnwidth,clip]{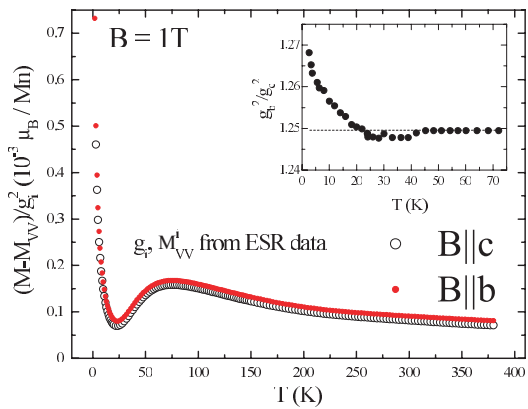}} \caption[] {\label{achsen}(color online) Magnetisation divided
  by the $g$-factor and corrected by anisotropic Van-Vleck
  magnetism along the respective axes of \sr\ for a constant
  magnetic field $B=1$~T versus temperature.}
\end{figure}

The fact that the free spins exhibit the similar anisotropy as
the dimerized ones is confirmed by our magnetization data. Our
analysis of $M(T)$ shows that the ratio of the Curie constants
which describe the response of the free spins along $b$- and
$c$-axis amounts to $C_b/C_c=g^2_b/g^2_c \simeq 1.25$. This
ratio agrees well with our ESR data. This fact is illustrated by
\figref{achsen} in which $M/g^2$ is shown for $B$ parallel to
the $b$-axis and the $c$-axis, respectively. After being
corrected for the anisotropy of the $g$ factor the magnetization
data for two orientations coincide almost perfectly in the whole
temperature range. Note that in our analysis we have also
considered the anisotropy of the Van-Vleck magnetism which is
related to the anisotropy of the $g$ factors as\cite{AbB:DP86}

\begin{eqnarray}
\chi_{c}^{VV} = \frac{8\cdot\mu_B^2}{\Delta_{yz,zx}} = (2-g_c)\frac{\mu_{\rm B}^2}{\lambda}
\\
\chi_{b}^{VV} = \frac{2\cdot\mu_B^2}{\Delta_{xy}}=
(2-g_b)\frac{\mu_{\rm B}^2}{\lambda}
\label{eqn.vv}
\ .
\end{eqnarray}

To summarize these results, our experimental data strongly
suggest that the free spins which cause the Curie-like response
in \sr\ are not extrinsic paramagnetic defects but reside in a
not perfectly dimerized chain. Thus our experimental study
qualitatively confirms the results of GL in
Ref.~\onlinecite{GeL:PRL04}, which predict the occurrence of
such free spins in the hole doped chain as a result of
structural modulations. As noted above, the theoretical
results suggest, however, a larger Curie-like response than
observed experimentally. In addition, while GL assume slightly
more than 60\% of holes in the chains of \sr , recent NEXAFS
experiments imply that there are less holes in the
chains,\cite{NMK:PRB00} i.e. there is a larger number of
magnetically active spins in the chains.

%===================    figure   =================================
%===================    figure   =================================

To resolve this contradiction we perform a numerical study to be
presented in the next section which gives evidence for trimers
in spin chains with less than 60\% hole doping. Trimers consist
of three next-nearest neighbor spins with a hole between each
spin, which are confined by two holes at each side, compare
\figref{F-1}. The presence of trimers might lift the discrepancy
between the calculated number of paramagnetic defects in the
dimerized state and the smaller experimental value of $N_S$. For
small external fields, trimers respond similarly to free spins.
The number of magnetically active spins involved in this
response is, however, three times larger. The first non-trivial
excitation of trimers occurs at $1.5J \sim 200$\,K, which can
not be observed in our $M(B)$ data at 2.5\,K because this is far
above the accessible field range. On the other hand, our $M(T)$
data do not rule out this scenario because at higher
temperatures there are indeed additional contributions to $M$
which are usually ascribed to the magnetism of the ladders and
to the melting of the charge order. Our numerical analysis in
section \ref{Exact} shows, that trimers (instead of or in
addition to single spins) perfectly explain our experimental
data. In contrast, the presence of quadrumers, which exhibit the
spin gap of $0.659J \sim 90$\,K is excluded by our $M(T)$ data
shown in \figref{sr14}(a). Our data also rule out that the
Curie-like response is due to nearest neighbor dimers, trimers,
etc., which are coupled ferromagnetically.

\section{Exact diagonalization studies\label{Exact}}

\begin{figure}[ht!]
\centering
\includegraphics[clip,width=35mm]{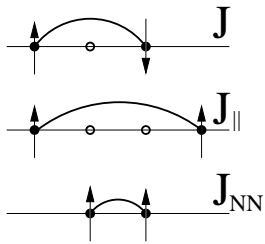}
\caption{Exchange parameters used for the exact diagonalization:
  $J=-134$~K, $J_\parallel=15$~K, and $J_{NN}=100$~K.
} \label{F-2}
\end{figure}

The dynamics of a hole-doped chain system constitutes a real
challenge for theoretical investigations. Especially the
evaluation of thermodynamic quantities both as function of
temperature and magnetic field is prohibitively complicated even
for moderate system sizes. In the following we are using an
effective spin Hamiltonian which depends parametrically on hole
positions.\cite{Sch:EPJB05} This ansatz is similar to a simple
Born-Oppenheimer description where the electronic Hamiltonian
(here spin Hamiltonian) depends parametrically on the positions
of the classical nuclei (here hole positions).  Each
configuration $\vec{c}$ of holes and spins defines a Hilbert
space which is orthogonal to all Hilbert spaces arising from
different configurations. The Hamilton operator
$\op{H}(\vec{c})$ of a certain configuration $\vec{c}$ is of
Heisenberg type, i.~e.
%--------------------------------------------------------
\begin{eqnarray}
\label{JS-1-1} \op{H} &=& \sum_{\vec{c}}\; \left( \op{H}(\vec{c}) + V(\vec{c}) \right)
\\
\label{JS-1-2} \op{H}(\vec{c}) &=& - \sum_{u<v}\; J_{uv}(\vec{c})\; \op{\vec{s}}(u) \cdot
\op{\vec{s}}(v)
\\
\label{JS-1-3} V(\vec{c}) &=& \frac{e^2}{4\pi\epsilon_0\,\epsilon_r\,r_0} \frac{1}{2}\; \sum_{u\ne
v}\; \frac{1}{|u-v|} \ .
\end{eqnarray}
%--------------------------------------------------------
$J_{uv}(\vec{c})$ are the respective exchange parameters which
depend on the configuration of holes. $J<0$ describes
antiferromagnetic coupling, $J>0$ ferromagnetic coupling.  For
the theoretical results presented in this article three exchange
parameters are used, see \figref{F-2}. The strongest and
antiferromagnetic exchange $J=-134$~K is across one hole. The
exchange accross two holes $J_\parallel=15$~K is ferromagnetic
as is the exchange $J_{NN}=100$~K of neighboring spins. Periodic
boundary conditions are applied for the following calculations,
i.e. large rings are considered instead of chains with open
boundaries.

%===================    figure   =================================
%===================    figure   =================================

It is obvious that different configurations of spins and holes
should also be energetically different. Besides their different
magnetic ground state energies resulting from \fmref{JS-1-2}
they also differ by their Coulomb interaction between the holes
and possibly by the Coulomb interaction with the local
environment. In the following we take the electrostatic
hole-hole repulsion by means of a screened Coulomb potential
into account. $r_0=2.75$~\AA\ is the distance between nearest
neighbor sites on the ring. The accurate value of the dielectric
constant $\epsilon_r$ is unknown. Several attempts have been
undertaken to estimate the dielectric constant which yielded
values for $\epsilon_r$ up to
30.\cite{CPP:PRL89,BaS:PRB94,EKZ:PRB97} If the interaction with
the local --
modulated\cite{IsT:JPSJ98,GeL:PRL04,GeL:EPJB05,GeL:05} --
environment is neglected, it is necessary to assume a rather
small dielectric constant $\epsilon_r\lessapprox 3$ in order to
simulate the magnetization data.\cite{Sch:EPJB05} Under such
conditions only the energetically lowest-lying spin-hole
configuration $\vec{c}$ contributes to the magnetization for
$T\lessapprox 200$~K.

%===================    figure   =================================
\begin{figure}[ht!]
\centering
\includegraphics[clip,width=45mm]{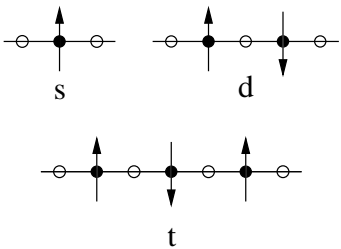}
\caption{Building blocks of the hole doped CuO$_2$ chains: open
  circles denote hole sites, arrows denote spin sites. The first
  block contains a single spin (s), the second a dimer (d), and
  the third block hosts three spins (t), which form a short
  chain.  }
\label{F-1}
\end{figure}
%===================    figure   =================================

%===================    figure   =================================
\begin{figure}[ht!]
\centering
\includegraphics[clip,width=0.8\columnwidth]{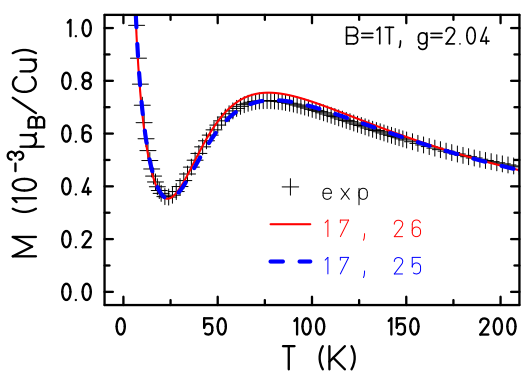}
\caption{(color online) Magnetization versus temperature for
  $B=1$~T: Experimental data ($\vec{B}\parallel$ $c$-axis) are
  given by crosses. The results of a complete numerical
  diagonalization are depicted by a solid curve for
  $N_{\text{s}}=17$ and $N_{\text{h}}=26$ as well as by a dashed
  curve for $N_{\text{s}}=17$ and $N_{\text{h}}=25$.}
\label{F-3}
\end{figure}
%===================    figure   =================================

The assumption of a perfectly dimerized chain was studied in
Ref.~\onlinecite{Sch:EPJB05}. In the following imperfect chains
will be investigated. To this end Hamiltonian \fmref{JS-1-1} is
numerically completely diagonalized for the currently largest
possible odd number of spins $N_{\text{s}}=17$. The number of
holes is chosen as $N_{\text{h}}=25$ or $N_{\text{h}}=26$ in
order to yield about 60~\% holes on the chain. Periodic boundary
conditions are applied. It turns out that the ground state
configurations are sequences of the building blocks shown in
\figref{F-1}. In the case of $N_{\text{s}}=17$ spins and
$N_{\text{h}}=26$ holes the resulting ground state sequence is
$\vec{c}_{17,26}=sdddddddd$, i.e. it contains only dimers except
for one single spin block.  This ground state configuration is
very similar to the one found using density functional theory
calculations.\cite{GeL:PRL04,GeL:05} In the case of
$N_{\text{s}}=17$ spins and $N_{\text{h}}=25$ holes the chain
consists of dimers and one block of three spins, i.e. it has the
sequence $\vec{c}_{17,25}=tddddddd$. In both cases other
configurations of spins and holes are energetically well
separated.

Having determined all energy eigenvalues and magnetic quantum
numbers, the magnetization can be evaluated for the two ground
state configurations. Figure \xref{F-3} shows the resulting
magnetization curves as a function of temperature for an applied
field of $B=1$~T. The solid curve displays the magnetization
curve for $N_{\text{s}}=17$ and $N_{\text{h}}=26$, i.e.
configuration $\vec{c}_{17,26}=sdddddddd$, whereas the dashed
curve shows the result for $N_{\text{s}}=17$ and
$N_{\text{h}}=25$, i.e. $\vec{c}_{17,25}=tddddddd$. Both curves
are rather close to the experimental data given by crosses.

\begin{figure}[ht!]
\centering
\includegraphics[clip,width=0.8\columnwidth]{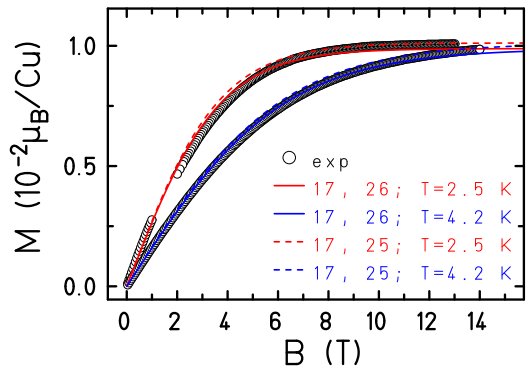}
\caption{(color online) Magnetization versus applied field for
  $T=2.5$~K and $T=4.2$~K: Experimental data ($\vec{B}\parallel$
  $c$-axis) are given by symbols. The results of a complete
  numerical diagonalization are depicted by a solid curves for
  $N_{\text{s}}=17$ and $N_{\text{h}}=26$ as well as by dashed
  curves for $N_{\text{s}}=17$ and $N_{\text{h}}=25$.}
\label{F-7}
\end{figure}

The magnetization has also been determined as a function of
applied field for two temperatures $T=2.5$~K and
$T=4.2$~K. Figure~\xref{F-7} shows that both configurations,
i.e. $\vec{c}_{17,26}=sdddddddd$ and $\vec{c}_{17,25}=tddddddd$
reproduce the data with high accuracy.

%===================    figure   =================================
%===================    figure   =================================

\section{Conclusion}

In this paper we present experimental as well as numerical
evidence that the Curie-like contribution to the low-temperature
magnetization of \sr\ is a genuine property of the CuO$_2$ chain
subsystem. The hole-doped chains are not regular sequences of
weekly interacting spin dimers, but instead they host a small
percentage of almost free spins and/or spin trimers. This
finding is in accord with recent density functional
calculations.\cite{GeL:PRL04,GeL:05} Nevertheless, our own
simulational studies favor the occurrence of trimers instead of
single spins which would be consitent with an intrinsic hole
doping of the chains of less than 60~\% as was inferred from
recent NEXAFS experiments.\cite{NMK:PRB00}

\begin{acknowledgments}
  This work was supported by the Deutsche Forschungsgemeinschaft
  (DFG) within SPP 1073 (BU887/1-3).  R.K. acknowledges support
  by the DFG through KL1824/1-1.
\end{acknowledgments}

%\bibliography{js-cup}

\begin{thebibliography}{30}
\expandafter\ifx\csname natexlab\endcsname\relax\def\natexlab#1{#1}\fi
\expandafter\ifx\csname bibnamefont\endcsname\relax
  \def\bibnamefont#1{#1}\fi
\expandafter\ifx\csname bibfnamefont\endcsname\relax
  \def\bibfnamefont#1{#1}\fi
\expandafter\ifx\csname citenamefont\endcsname\relax
  \def\citenamefont#1{#1}\fi
\expandafter\ifx\csname url\endcsname\relax
  \def\url#1{\texttt{#1}}\fi
\expandafter\ifx\csname urlprefix\endcsname\relax\def\urlprefix{URL }\fi
\providecommand{\bibinfo}[2]{#2}
\providecommand{\eprint}[2][]{\url{#2}}

\bibitem[{\citenamefont{Kivelson et~al.}(2003)\citenamefont{Kivelson, Bindloss,
  Fradkin, Oganesyan, Tranquada, Kapitulnik, and Howald}}]{KBF:RMP03}
\bibinfo{author}{\bibfnamefont{S.~A.} \bibnamefont{Kivelson}},
  \bibinfo{author}{\bibfnamefont{I.~P.} \bibnamefont{Bindloss}},
  \bibinfo{author}{\bibfnamefont{E.}~\bibnamefont{Fradkin}},
  \bibinfo{author}{\bibfnamefont{V.}~\bibnamefont{Oganesyan}},
  \bibinfo{author}{\bibfnamefont{J.~M.} \bibnamefont{Tranquada}},
  \bibinfo{author}{\bibfnamefont{A.}~\bibnamefont{Kapitulnik}},
  \bibnamefont{and} \bibinfo{author}{\bibfnamefont{C.}~\bibnamefont{Howald}},
  \bibinfo{journal}{Rev. Mod. Phys.} \textbf{\bibinfo{volume}{75}},
  \bibinfo{pages}{1201} (\bibinfo{year}{2003}).

\bibitem[{\citenamefont{Uehara et~al.}(1996)\citenamefont{Uehara, Nagata,
  Akimitsu, Takahashi, Mori, and Kinoshita}}]{UNA:JPSJ96}
\bibinfo{author}{\bibfnamefont{M.}~\bibnamefont{Uehara}},
  \bibinfo{author}{\bibfnamefont{T.}~\bibnamefont{Nagata}},
  \bibinfo{author}{\bibfnamefont{J.}~\bibnamefont{Akimitsu}},
  \bibinfo{author}{\bibfnamefont{H.}~\bibnamefont{Takahashi}},
  \bibinfo{author}{\bibfnamefont{N.}~\bibnamefont{Mori}}, \bibnamefont{and}
  \bibinfo{author}{\bibfnamefont{K.}~\bibnamefont{Kinoshita}},
  \bibinfo{journal}{J. Phys. Soc. Jpn.} \textbf{\bibinfo{volume}{65}},
  \bibinfo{pages}{2764} (\bibinfo{year}{1996}).

\bibitem[{\citenamefont{Dagotto}(1999)}]{Dag:RPP99}
\bibinfo{author}{\bibfnamefont{E.}~\bibnamefont{Dagotto}},
  \bibinfo{journal}{Rep. Prog. Phys.} \textbf{\bibinfo{volume}{62}},
  \bibinfo{pages}{1525} (\bibinfo{year}{1999}).

\bibitem[{\citenamefont{Horsch et~al.}(2005)\citenamefont{Horsch, Sofin, Mayr,
  and Jansen}}]{HSM:PRL05}
\bibinfo{author}{\bibfnamefont{P.}~\bibnamefont{Horsch}},
  \bibinfo{author}{\bibfnamefont{M.}~\bibnamefont{Sofin}},
  \bibinfo{author}{\bibfnamefont{M.}~\bibnamefont{Mayr}}, \bibnamefont{and}
  \bibinfo{author}{\bibfnamefont{M.}~\bibnamefont{Jansen}},
  \bibinfo{journal}{Phys. Rev. Lett.} \textbf{\bibinfo{volume}{94}},
  \bibinfo{pages}{076403} (\bibinfo{year}{2005}).

\bibitem[{\citenamefont{Matsuda et~al.}(2005)\citenamefont{Matsuda, Kakurai,
  Kurogi, Kudo, Koike, Yamaguchi, Ito, and Oka}}]{MKK:PRB05}
\bibinfo{author}{\bibfnamefont{M.}~\bibnamefont{Matsuda}},
  \bibinfo{author}{\bibfnamefont{K.}~\bibnamefont{Kakurai}},
  \bibinfo{author}{\bibfnamefont{S.}~\bibnamefont{Kurogi}},
  \bibinfo{author}{\bibfnamefont{K.}~\bibnamefont{Kudo}},
  \bibinfo{author}{\bibfnamefont{Y.}~\bibnamefont{Koike}},
  \bibinfo{author}{\bibfnamefont{H.}~\bibnamefont{Yamaguchi}},
  \bibinfo{author}{\bibfnamefont{T.}~\bibnamefont{Ito}}, \bibnamefont{and}
  \bibinfo{author}{\bibfnamefont{K.}~\bibnamefont{Oka}},
  \bibinfo{journal}{Phys. Rev. B} \textbf{\bibinfo{volume}{71}},
  \bibinfo{pages}{104414} (\bibinfo{year}{2005}).

\bibitem[{\citenamefont{Tsuji et~al.}(2005)\citenamefont{Tsuji, Kumagai, Kato, and Koike}}]{TKK:JPSJ96}
\bibinfo{author}{\bibfnamefont{S.}~\bibnamefont{Tsuji}},
  \bibinfo{author}{\bibfnamefont{K.}~\bibnamefont{Kumagai}},
  \bibinfo{author}{\bibfnamefont{M.}~\bibnamefont{Kato}}, \bibnamefont{and}
  \bibinfo{author}{\bibfnamefont{Y.}~\bibnamefont{Koike}},
  \bibinfo{journal}{J. Phys. Soc. Jpn.} \textbf{\bibinfo{volume}{65}},
  \bibinfo{pages}{3474} (\bibinfo{year}{1996}).

\bibitem[{\citenamefont{Eccleston et~al.}(1998)\citenamefont{Eccleston, Uehara,
  Akimitsu, Eisaki, Motoyama, and Uchida}}]{EUA:PRL98}
\bibinfo{author}{\bibfnamefont{R.~S.} \bibnamefont{Eccleston}},
  \bibinfo{author}{\bibfnamefont{M.}~\bibnamefont{Uehara}},
  \bibinfo{author}{\bibfnamefont{J.}~\bibnamefont{Akimitsu}},
  \bibinfo{author}{\bibfnamefont{H.}~\bibnamefont{Eisaki}},
  \bibinfo{author}{\bibfnamefont{N.}~\bibnamefont{Motoyama}}, \bibnamefont{and}
  \bibinfo{author}{\bibfnamefont{S. I.}~\bibnamefont{Uchida}},
  \bibinfo{journal}{Phys. Rev. Lett.} \textbf{\bibinfo{volume}{81}},
  \bibinfo{pages}{1702} (\bibinfo{year}{1998}).

\bibitem[{\citenamefont{Katano et~al.}(1999)\citenamefont{Katano, Nagata,
  Akimitsu, Nishi, and Kakurai}}]{KNA:PRL99}
\bibinfo{author}{\bibfnamefont{S.}~\bibnamefont{Katano}},
  \bibinfo{author}{\bibfnamefont{T.}~\bibnamefont{Nagata}},
  \bibinfo{author}{\bibfnamefont{J.}~\bibnamefont{Akimitsu}},
  \bibinfo{author}{\bibfnamefont{M.}~\bibnamefont{Nishi}}, \bibnamefont{and}
  \bibinfo{author}{\bibfnamefont{K.}~\bibnamefont{Kakurai}},
  \bibinfo{journal}{Phys. Rev. Lett.} \textbf{\bibinfo{volume}{82}},
  \bibinfo{pages}{636} (\bibinfo{year}{1999}).

\bibitem[{\citenamefont{Hess et~al.}(2001)\citenamefont{Hess, Baumann,
  Ammerahl, B\"uchner, Heidrich-Meisner, Brenig, and Revcolevschi}}]{HBA:PRB01}
\bibinfo{author}{\bibfnamefont{C.}~\bibnamefont{Hess}},
  \bibinfo{author}{\bibfnamefont{C.}~\bibnamefont{Baumann}},
  \bibinfo{author}{\bibfnamefont{U.}~\bibnamefont{Ammerahl}},
  \bibinfo{author}{\bibfnamefont{B.}~\bibnamefont{B\"uchner}},
  \bibinfo{author}{\bibfnamefont{F.}~\bibnamefont{Heidrich-Meisner}},
  \bibinfo{author}{\bibfnamefont{W.}~\bibnamefont{Brenig}}, \bibnamefont{and}
  \bibinfo{author}{\bibfnamefont{A.}~\bibnamefont{Revcolevschi}},
  \bibinfo{journal}{Phys. Rev. B} \textbf{\bibinfo{volume}{64}},
  \bibinfo{pages}{184305} (\bibinfo{year}{2001}).

%\bibitem[{\citenamefont{Kato et~al.}(1996)\citenamefont{Kato, Shiota, and Koike}}]{KSK:PC96}
%\bibinfo{author}{\bibfnamefont{M.}~\bibnamefont{Kato}},
%\bibinfo{author}{\bibfnamefont{K.}~\bibnamefont{Shiota}}, \bibnamefont{and}
%  \bibinfo{author}{\bibfnamefont{Y.}~\bibnamefont{Koike}},
%  \bibinfo{journal}{Physica C} \textbf{\bibinfo{volume}{258}},
%  \bibinfo{pages}{284} (\bibinfo{year}{1996}).

\bibitem{KSK:PC96}
M.\ Kato, K.\ Shiota, and Y.\ Koike, Physica\ C {\bf 258}, 284 (1996); M.\ Kato, T.\ Adachi, and Y.\ Koike, Physica C {\bf 265}, 107 (1996).

\bibitem[{\citenamefont{Eccleston et~al.}(1996)\citenamefont{Eccleston, Azuma,
  and Takano}}]{EAT:PRB96}
\bibinfo{author}{\bibfnamefont{R.~S.} \bibnamefont{Eccleston}},
  \bibinfo{author}{\bibfnamefont{M.}~\bibnamefont{Azuma}}, \bibnamefont{and}
  \bibinfo{author}{\bibfnamefont{M.}~\bibnamefont{Takano}},
  \bibinfo{journal}{Phys. Rev. B} \textbf{\bibinfo{volume}{53}},
  \bibinfo{pages}{R14721} (\bibinfo{year}{1996}).

\bibitem[{\citenamefont{Matsuda et~al.}(1996)\citenamefont{Matsuda, Katsumata,
  Yokoo, Shapiro, and Shirane}}]{MKY:PRB96}
\bibinfo{author}{\bibfnamefont{M.}~\bibnamefont{Matsuda}},
  \bibinfo{author}{\bibfnamefont{K.}~\bibnamefont{Katsumata}},
  \bibinfo{author}{\bibfnamefont{T.}~\bibnamefont{Yokoo}},
  \bibinfo{author}{\bibfnamefont{S.~M.} \bibnamefont{Shapiro}},
  \bibnamefont{and} \bibinfo{author}{\bibfnamefont{G.}~\bibnamefont{Shirane}},
  \bibinfo{journal}{Phys. Rev. B} \textbf{\bibinfo{volume}{54}},
  \bibinfo{pages}{R15626} (\bibinfo{year}{1996}).

\bibitem[{\citenamefont{Regnault et~al.}(1999)\citenamefont{Regnault, Boucher,
  Moudden, Lorenzo, Hiess, Ammerahl, Dhalenne, and Revcolevschi}}]{RBM:PRB99}
\bibinfo{author}{\bibfnamefont{L.~P.} \bibnamefont{Regnault}},
  \bibinfo{author}{\bibfnamefont{J.~P.} \bibnamefont{Boucher}},
  \bibinfo{author}{\bibfnamefont{H.}~\bibnamefont{Moudden}},
  \bibinfo{author}{\bibfnamefont{J.~E.} \bibnamefont{Lorenzo}},
  \bibinfo{author}{\bibfnamefont{A.}~\bibnamefont{Hiess}},
  \bibinfo{author}{\bibfnamefont{U.}~\bibnamefont{Ammerahl}},
  \bibinfo{author}{\bibfnamefont{G.}~\bibnamefont{Dhalenne}}, \bibnamefont{and}
  \bibinfo{author}{\bibfnamefont{A.}~\bibnamefont{Revcolevschi}},
  \bibinfo{journal}{Phys. Rev. B} \textbf{\bibinfo{volume}{59}},
  \bibinfo{pages}{1055} (\bibinfo{year}{1999}).

\bibitem[{\citenamefont{Matsuda et~al.}(1999)\citenamefont{Matsuda, Yosihama,
  Kakurai, and Shirane}}]{MYK:PRB99}
\bibinfo{author}{\bibfnamefont{M.}~\bibnamefont{Matsuda}},
  \bibinfo{author}{\bibfnamefont{T.}~\bibnamefont{Yosihama}},
  \bibinfo{author}{\bibfnamefont{K.}~\bibnamefont{Kakurai}}, \bibnamefont{and}
  \bibinfo{author}{\bibfnamefont{G.}~\bibnamefont{Shirane}},
  \bibinfo{journal}{Phys. Rev. B} \textbf{\bibinfo{volume}{59}},
  \bibinfo{pages}{1060} (\bibinfo{year}{1999}).

\bibitem[{\citenamefont{Takigawa et~al.}(1998)\citenamefont{Takigawa, Motoyama,
  Eisaki, and Uchida}}]{TME:PRB98}
\bibinfo{author}{\bibfnamefont{M.}~\bibnamefont{Takigawa}},
  \bibinfo{author}{\bibfnamefont{N.}~\bibnamefont{Motoyama}},
  \bibinfo{author}{\bibfnamefont{H.}~\bibnamefont{Eisaki}}, \bibnamefont{and}
  \bibinfo{author}{\bibfnamefont{S.}~\bibnamefont{Uchida}},
  \bibinfo{journal}{Phys. Rev. B} \textbf{\bibinfo{volume}{57}},
  \bibinfo{pages}{1124} (\bibinfo{year}{1998}).

\bibitem[{\citenamefont{Kataev et~al.}(2001{\natexlab{a}})\citenamefont{Kataev,
  Choi, Gr\"uninger, Ammerahl, B\"uchner, Freimuth, and
  Revcolevschi}}]{KCG:PRB01}
\bibinfo{author}{\bibfnamefont{V.}~\bibnamefont{Kataev}},
  \bibinfo{author}{\bibfnamefont{K.~Y.} \bibnamefont{Choi}},
  \bibinfo{author}{\bibfnamefont{M.}~\bibnamefont{Gr\"uninger}},
  \bibinfo{author}{\bibfnamefont{U.}~\bibnamefont{Ammerahl}},
  \bibinfo{author}{\bibfnamefont{B.}~\bibnamefont{B\"uchner}},
  \bibinfo{author}{\bibfnamefont{A.}~\bibnamefont{Freimuth}}, \bibnamefont{and}
  \bibinfo{author}{\bibfnamefont{A.}~\bibnamefont{Revcolevschi}},
  \bibinfo{journal}{Phys. Rev. B} \textbf{\bibinfo{volume}{64}},
  \bibinfo{pages}{104422} (\bibinfo{year}{2001}{\natexlab{a}}).

\bibitem[{\citenamefont{Gelle and Lepetit}(2004{\natexlab{a}})}]{GeL:PRL04}
\bibinfo{author}{\bibfnamefont{A.}~\bibnamefont{Gelle}} \bibnamefont{and}
  \bibinfo{author}{\bibfnamefont{M.~B.} \bibnamefont{Lepetit}},
  \bibinfo{journal}{Phys. Rev. Lett.} \textbf{\bibinfo{volume}{92}},
  \bibinfo{pages}{236402} (\bibinfo{year}{2004}{\natexlab{a}}).

\bibitem[{\citenamefont{Schnack and Ouchni}(2005)}]{ScO:JMMM05}
\bibinfo{author}{\bibfnamefont{J.}~\bibnamefont{Schnack}} \bibnamefont{and}
  \bibinfo{author}{\bibfnamefont{F.}~\bibnamefont{Ouchni}},
  \bibinfo{journal}{J. Magn. Magn. Mater.} \textbf{\bibinfo{volume}{290-291}},
  \bibinfo{pages}{341} (\bibinfo{year}{2005}).

\bibitem[{\citenamefont{Schnack}(2005)}]{Sch:EPJB05}
\bibinfo{author}{\bibfnamefont{J.}~\bibnamefont{Schnack}},
  \bibinfo{journal}{Eur. Phys. J. B} \textbf{\bibinfo{volume}{45}},
  \bibinfo{pages}{311} (\bibinfo{year}{2005}).

\bibitem[{\citenamefont{Ammerahl and Revcolevschi}(1999)}]{AmR:JCG99}
\bibinfo{author}{\bibfnamefont{U.}~\bibnamefont{Ammerahl}} \bibnamefont{and}
  \bibinfo{author}{\bibfnamefont{A.}~\bibnamefont{Revcolevschi}},
  \bibinfo{journal}{J. Cryst. Growth} \textbf{\bibinfo{volume}{197}},
  \bibinfo{pages}{825} (\bibinfo{year}{1999}).

\bibitem[{\citenamefont{Ammerahl et~al.}(2000)\citenamefont{Ammerahl,
  B\"uchner, Colonescu, Gross, and Revcolevschi}}]{ABC:PRB00}
\bibinfo{author}{\bibfnamefont{U.}~\bibnamefont{Ammerahl}},
  \bibinfo{author}{\bibfnamefont{B.}~\bibnamefont{B\"uchner}},
  \bibinfo{author}{\bibfnamefont{L.}~\bibnamefont{Colonescu}},
  \bibinfo{author}{\bibfnamefont{R.}~\bibnamefont{Gross}}, \bibnamefont{and}
  \bibinfo{author}{\bibfnamefont{A.}~\bibnamefont{Revcolevschi}},
  \bibinfo{journal}{Phys. Rev. B} \textbf{\bibinfo{volume}{62}},
  \bibinfo{pages}{8630} (\bibinfo{year}{2000}).

\bibitem[{\citenamefont{Hess et~al.}(2004)\citenamefont{Hess, {ElHaes},
  B\"uchner, Ammerahl, H\"ucker, and Revcolevschi}}]{HEB:PRL04}
\bibinfo{author}{\bibfnamefont{C.}~\bibnamefont{Hess}},
  \bibinfo{author}{\bibfnamefont{H.}~\bibnamefont{ElHaes}},
  \bibinfo{author}{\bibfnamefont{B.}~\bibnamefont{B\"uchner}},
  \bibinfo{author}{\bibfnamefont{U.}~\bibnamefont{Ammerahl}},
  \bibinfo{author}{\bibfnamefont{M.}~\bibnamefont{H\"ucker}}, \bibnamefont{and}
  \bibinfo{author}{\bibfnamefont{A.}~\bibnamefont{Revcolevschi}},
  \bibinfo{journal}{Phys. Rev. Lett.} \textbf{\bibinfo{volume}{93}},
  \bibinfo{pages}{027005} (\bibinfo{year}{2004}).

\bibitem[{\citenamefont{Carter et~al.}(1996)\citenamefont{Carter, Batlogg,
  Cava, Krajewski, Peck, and Rice}}]{CBC:PRL96}
\bibinfo{author}{\bibfnamefont{S.~A.} \bibnamefont{Carter}},
  \bibinfo{author}{\bibfnamefont{B.}~\bibnamefont{Batlogg}},
  \bibinfo{author}{\bibfnamefont{R.~J.} \bibnamefont{Cava}},
  \bibinfo{author}{\bibfnamefont{J.~J.} \bibnamefont{Krajewski}},
  \bibinfo{author}{\bibfnamefont{W.~F.} \bibnamefont{Peck}}, \bibnamefont{and}
  \bibinfo{author}{\bibfnamefont{T.~M.} \bibnamefont{Rice}},
  \bibinfo{journal}{Phys. Rev. Lett.} \textbf{\bibinfo{volume}{77}},
  \bibinfo{pages}{1378} (\bibinfo{year}{1996}).


\bibitem{comment}
A possible intsinsic nature of the Curie contribution to the chain susceptibilty $\chi_{chain}$ was first pointed out in Ref.~\onlinecite{KSK:PC96}
from the analysis of the $T$ dependence of $\chi_{chain}$ of polycrystalline samples.



\bibitem[{\citenamefont{Kataev et~al.}(2001{\natexlab{b}})\citenamefont{Kataev,
  Choi, Gr\"uninger, Ammerahl, B\"uchner, Freimuth, and
  Revcolevschi}}]{KCG:PRL01}
\bibinfo{author}{\bibfnamefont{V.}~\bibnamefont{Kataev}},
  \bibinfo{author}{\bibfnamefont{K.~Y.} \bibnamefont{Choi}},
  \bibinfo{author}{\bibfnamefont{M.}~\bibnamefont{Gr\"uninger}},
  \bibinfo{author}{\bibfnamefont{U.}~\bibnamefont{Ammerahl}},
  \bibinfo{author}{\bibfnamefont{B.}~\bibnamefont{B\"uchner}},
  \bibinfo{author}{\bibfnamefont{A.}~\bibnamefont{Freimuth}}, \bibnamefont{and}
  \bibinfo{author}{\bibfnamefont{A.}~\bibnamefont{Revcolevschi}},
  \bibinfo{journal}{Phys. Rev. Lett.} \textbf{\bibinfo{volume}{86}},
  \bibinfo{pages}{2882} (\bibinfo{year}{2001}{\natexlab{b}}).

\bibitem[{\citenamefont{Abragam and Bleaney}(1986)}]{AbB:DP86}
\bibinfo{author}{\bibfnamefont{A.}~\bibnamefont{Abragam}} \bibnamefont{and}
  \bibinfo{author}{\bibfnamefont{B.}~\bibnamefont{Bleaney}},
  \emph{\bibinfo{title}{Electron Paramagnetic Resonance of Transition Ions}}
  (\bibinfo{publisher}{Dover Publications}, \bibinfo{address}{New York},
  \bibinfo{year}{1986}).

\bibitem[{\citenamefont{Chen et~al.}(1989)\citenamefont{Chen, Preyer, Picone,
  Kastner, Jenssen, Gabbe, Cassanho, and Birgeneau}}]{CPP:PRL89}
\bibinfo{author}{\bibfnamefont{C.~Y.} \bibnamefont{Chen}},
  \bibinfo{author}{\bibfnamefont{N.~W.} \bibnamefont{Preyer}},
  \bibinfo{author}{\bibfnamefont{P.~J.} \bibnamefont{Picone}},
  \bibinfo{author}{\bibfnamefont{M.~A.} \bibnamefont{Kastner}},
  \bibinfo{author}{\bibfnamefont{H.~P.} \bibnamefont{Jenssen}},
  \bibinfo{author}{\bibfnamefont{D.~R.} \bibnamefont{Gabbe}},
  \bibinfo{author}{\bibfnamefont{A.}~\bibnamefont{Cassanho}}, \bibnamefont{and}
  \bibinfo{author}{\bibfnamefont{R.~J.} \bibnamefont{Birgeneau}},
  \bibinfo{journal}{Phys. Rev. Lett.} \textbf{\bibinfo{volume}{63}},
  \bibinfo{pages}{2307} (\bibinfo{year}{1989}).

\bibitem[{\citenamefont{N\"ucker et~al.}(2000)\citenamefont{N\"ucker, Merz,
  Kuntscher, Gerhold, Schuppler, Neudert, Golden, Fink, Schild, Stadler
  et~al.}}]{NMK:PRB00}
\bibinfo{author}{\bibfnamefont{N.}~\bibnamefont{N\"ucker}},
  \bibinfo{author}{\bibfnamefont{M.}~\bibnamefont{Merz}},
  \bibinfo{author}{\bibfnamefont{C.~A.} \bibnamefont{Kuntscher}},
  \bibinfo{author}{\bibfnamefont{S.}~\bibnamefont{Gerhold}},
  \bibinfo{author}{\bibfnamefont{S.}~\bibnamefont{Schuppler}},
  \bibinfo{author}{\bibfnamefont{R.}~\bibnamefont{Neudert}},
  \bibinfo{author}{\bibfnamefont{M.~S.} \bibnamefont{Golden}},
  \bibinfo{author}{\bibfnamefont{J.}~\bibnamefont{Fink}},
  \bibinfo{author}{\bibfnamefont{D.}~\bibnamefont{Schild}},
  \bibinfo{author}{\bibfnamefont{S.}~\bibnamefont{Stadler}},
  \bibnamefont{et~al.}, \bibinfo{journal}{Phys. Rev. B}
  \textbf{\bibinfo{volume}{62}}, \bibinfo{pages}{14384} (\bibinfo{year}{2000}).

\bibitem[{\citenamefont{Barriquand and Sawatzky}(1994)}]{BaS:PRB94}
\bibinfo{author}{\bibfnamefont{F.}~\bibnamefont{Barriquand}} \bibnamefont{and}
  \bibinfo{author}{\bibfnamefont{G.~A.} \bibnamefont{Sawatzky}},
  \bibinfo{journal}{Phys. Rev. B} \textbf{\bibinfo{volume}{50}},
  \bibinfo{pages}{16649} (\bibinfo{year}{1994}).

\bibitem[{\citenamefont{Emery et~al.}(1997)\citenamefont{Emery, Kivelson, and
  Zachar}}]{EKZ:PRB97}
\bibinfo{author}{\bibfnamefont{V.~J.} \bibnamefont{Emery}},
  \bibinfo{author}{\bibfnamefont{S.~A.} \bibnamefont{Kivelson}},
  \bibnamefont{and} \bibinfo{author}{\bibfnamefont{O.}~\bibnamefont{Zachar}},
  \bibinfo{journal}{Phys. Rev. B} \textbf{\bibinfo{volume}{56}},
  \bibinfo{pages}{6120} (\bibinfo{year}{1997}).

\bibitem[{\citenamefont{Isobe and Takayama-Muromachi}(1998)}]{IsT:JPSJ98}
\bibinfo{author}{\bibfnamefont{M.}~\bibnamefont{Isobe}} \bibnamefont{and}
  \bibinfo{author}{\bibfnamefont{E.}~\bibnamefont{Takayama-Muromachi}},
  \bibinfo{journal}{J. Phys. Soc. Jpn.} \textbf{\bibinfo{volume}{67}},
  \bibinfo{pages}{3119} (\bibinfo{year}{1998}).

\bibitem[{\citenamefont{Gelle and Lepetit}(2005)}]{GeL:EPJB05}
\bibinfo{author}{\bibfnamefont{A.}~\bibnamefont{Gelle}} \bibnamefont{and}
  \bibinfo{author}{\bibfnamefont{M.~B.} \bibnamefont{Lepetit}},
  \bibinfo{journal}{Eur. Phys. J. B} \textbf{\bibinfo{volume}{43}},
  \bibinfo{pages}{29} (\bibinfo{year}{2005}).

\bibitem[{\citenamefont{Gelle and Lepetit}(2004{\natexlab{b}})}]{GeL:05}
\bibinfo{author}{\bibfnamefont{A.}~\bibnamefont{Gelle}} \bibnamefont{and}
  \bibinfo{author}{\bibfnamefont{M.~B.} \bibnamefont{Lepetit}}
  (\bibinfo{year}{2004}{\natexlab{b}}), \bibinfo{note}{unpublished,
  cond-mat/0410203}.

\end{thebibliography}

\end{document}